\PassOptionsToPackage{table}{xcolor}
\documentclass[sigconf,balance=false,letterpaper]{acmart}
\setcitestyle{square}

\setcopyright{rightsretained} 
\copyrightyear{2022} 
\acmYear{2022} 
\acmConference{SIGGRAPH '22 Talks}{August 07-11, 2022}{Vancouver, BC, Canada}\acmBooktitle{Special Interest Group on Computer Graphics and Interactive Techniques Conference Talks (SIGGRAPH '22 Talks), August 07-11, 2022}\acmDOI{10.1145/3532836.3536227}
\acmISBN{978-1-4503-9371-3/22/08}

\usepackage[mathletters]{ucs}
\usepackage[utf8x]{inputenc}

\newcommand{\paragraphtight}[1]{\paragraph{#1}}

\renewcommand{\binary}{O1}

\newcommand{\var}{O4}
\newcommand{\dataset}{O5}
\newcommand{\userstudy}{O6}
\newcommand{\classifier}{O7}

\usepackage{tabularx}
\usepackage{booktabs}
\usepackage{makecell}
\usepackage[table]{xcolor}
\definecolor{white}{rgb}{1,1,1}
\definecolor{lightbluishgrey}{rgb}{0.76471,0.84824,0.91647}

\usepackage{subcaption}
\usepackage{wrapfig}
\usepackage{graphicx}

\usepackage{listings}
\usepackage{layouts}
\makeatletter
\newcommand{\layoutdetails}{%
\begin{tabular}{ll}
 \texttt{\textbackslash{textwidth}} & \printinunitsof{in}\prntlen{\textwidth} \\
\texttt{\textbackslash{linewidth}} & \printinunitsof{in}\prntlen{\linewidth} \\
Main text font &  \f@size pt \f@family \\
\sffamily \small Caption text font &  \sffamily \small \f@size pt \f@family \\
\end{tabular}%
}

\newcommand{\PreserveBackslash}[1]{\let\temp=\\#1\let\\=\temp}
\newcolumntype{C}[1]{>{\PreserveBackslash\centering}p{#1}}
\makeatother

\begin{document}

\title{Sex and Gender in the Computer Graphics Research Literature}

\author{Ana Dodik}\authornote{Joint First Authors}\authornote{Work done while the author was employed by Meta Platforms, Inc.}
\affiliation{\institution{MIT}\country{United States of America}}
\email{anadodik@mit.edu}

\author{Silvia Sellán}\authornotemark[1]
\affiliation{\institution{University of Toronto}\country{Canada}}
\email{sgsellan@cs.toronto.edu}

\author{Theodore Kim}
\affiliation{\institution{Yale University}\country{United States of America}}\email{theodore.kim@yale.edu}

\author{Amanda Phillips}
\affiliation{\institution{Georgetown University}\country{United States of America}}
\email{adp77@georgetown.edu}

\maketitle

\section{Introduction}

Sex and gender are referenced in the Computer Graphics literature: a dataset is said to contain images of men and women, user study participants are reported with certain male/female ratios, a body modeling algorithm trains two different gendered models, etc.

The scientific consensus around sex and gender has evolved in the past decades \cite{pmid30377332}. As surveyed by \citet{fausto2012sex}, \emph{sex} is not one but a combination of many biological classifications (\emph{chromosomal, hormonal, reproductive, ...}) which cannot be assigned in a binary way to one in 50 people \cite{blackless2000sexually}.
\emph{Gender} refers to an individual's self-identity \cite{money1972man}, their performance as shaped by social expectations \cite{butler2003gender}, or organizational structures that segregate people into, e.g. different bathrooms or professions \cite{lorber1994paradoxes}. In these contemporary definitions, gender is fluid, culturally-specific and not binary. Assuming outdated binary definitions of sex and gender is not just scientifically incorrect, but also harmful to those who conform the least to this binary (e.g, \emph{intersex}, \emph{transgender}, \emph{non-binary} people), whom we call \emph{gender non-conforming} \cite{un2015report}.

The treatment of sex and gender in SIGGRAPH Technical Papers
still adheres to a binary understanding, excluding gender non-conforming people. Further, it makes research lag behind the needs of industry. The latest character modeller for \citet{metahuman} and the Cloud Vision API by \citet{googlegender} have removed references to sex and gender. \emph{Animal Crossing} and \emph{Forza Horizon} decouple attributes like body proportions, voice pitch and pronouns. 

We will use an algorithmic fairness lens to argue that this binary understanding adds algorithmic biases detrimental to scientific integrity.
We will examine the real-world harms caused these biases in how gender non-conforming people interact with
our technology. We advocate for a reexamination of our treatment of gender, and show that correcting problematic practices in
our community will open the door to new avenues of research.

\section{Survey}

Inspired by \citet{keyes2018misgendering}, we survey
all technical papers presented at any SIGGRAPH since 2015. We list all 64 containing mentions of sex or gender in our \textbf{Supplemental Material}, along with our main observations (O1-7). We make the deliberate choice to reference these observations and not specific works in this main text to stress that we do not associate any malicious intent to individual authors. Rather, we are showing how seemingly neutral, well-established practices in our community (which includes journals, editors, reviewers, etc.) can unwittingly perpetuate forms of algorithmic bias.

Our observed references to sex and gender varied in nature from demographic information regarding study participants (\userstudy) or dataset makeup (\dataset) to gender-specific algorithms (\var). Whenever gender or sex is used
explicitly as a variable, it is always a binary (\binary) proxy for features such as body proportions or speech characteristics. The existence of gender non-conforming people was never acknowledged (O3). We found works proposing or using image-based (binary) gender recognition algorithms (\classifier).

\subsection{Algorithmic Fairness Analysis}\label{sec:analysis}

Our survey shows that the current use of gender and sex in Computer Graphics is at best ill-defined, and at worst incorrect.
We apply the framework of \citet{Suresh2021}, which categorizes bias by its stage in a system's lifecycle
(cf. \cite{fairnesssurvey,FriedmanAndNissenbaum,olteanu2019social}).
We give examples of how different biases occur, showing them to be \emph{technical} limitations that impede the development of precise, reproducible research.

\paragraphtight{Representation bias} Portions of populations may be poorly represented by a dataset, either because of \emph{sample selection bias} (e.g. a sampling procedure which excludes non-binary people) or due to the use of uniform sampling which would poorly represent sex and gender minorities. Despite the prevalence of these individuals in the general population, not a single paper (O3) explicitly mentioned them as part of datasets (\dataset) or user study participants (\userstudy). 
This may be due to measurement bias or an accidentally exclusionary sampling procedure. No works analyzed any type of representation bias experienced by gender non-conforming individuals (\binary).

\paragraphtight{Historical bias} Data, despite being abundant and perfectly sampled, may encode existing prejudice. For example, a \emph{gender classifier} (\classifier) trained on portrait image data collected in an environment where social norms dictate gender expression might learn that ``wearing a dress'' means woman, and ``short hair'' means man.

\paragraphtight{Measurement bias} Bias may be introduced through the selection and measurement of features and target variables.
Many works use sex or gender as imprecise \emph{proxies} (\var) for attributes like \emph{commonly co-occuring bodily} or \emph{speech characteristics}, in lieu of less abstract features like hair length or voice pitch. Some works even combined proxies, e.g., conversational agents that use gender for voice pitch \emph{and} culturally acquired speech inflections.

When gender or sex was chosen as a variable, it was always (\binary) through an \emph{inaccurate method of measurement}, treating them as binary variables that exclude gender and sex minorities by design. Alternatively, (\classifier) \emph{incorrect methods of measurement} were used, such as image-based gender classifiers in lieu of self-identification, which can misidentify gender non-conforming individuals.

\paragraphtight{Omitted variable bias} A successful feature may correlate with an important feature that has been omitted from the model (see e.g., \cite{clarke2005phantom}). \emph{Gender} or \emph{sex} are likely not as discriminative when the result is also conditioned on \emph{hair length}, \emph{hip width} or \emph{mean voice frequency}. When the use of gender or sex was justified because of an assumed improvement in accuracy (\var), we found no attempt to identify if the success was due to omitted variables.

\paragraphtight{Evaluation bias} These are biases occur during evaluation of an algorithm, such as body modeling works that provide binary segregated parametric models (O3). These are then used to evaluate \emph{other} works with orthogonal contributions, like virtual try-on or motion capture. If our community codifies biased benchmarks, we encourage the development of models that conform to those biases.

\paragraphtight{Deployment bias} Real-world harm is introduced when graphics models are published or deployed. The exclusive publication of papers with a binary understanding of sex and gender incentivizes researchers (and reviewers) to conform to that definition (\binary).
This leads to \emph{feedback loops}: if gender non-conforming people are not included in a virtual clothing try-on system, they are less likely to use it, skewing the system's performance data to include them even less. Finally, a system can impose its biases onto user behavior: a trans person may need to change the pitch of their voice in order to not get misgendered by an algorithm, further skewing the data.

\subsection{Real world harm}

The technical limitations of the reviewed algorithms can lead to real world harms. As Computer Graphics is increasingly applied to other fields, such as geometric data processing in medicine, or for synthetic dataset generation in computer vision, with numerous downstream applications \cite{cars, chen2021synthetic, dhs}, it is paramount to understand that our algorithms can and will be used in novel ways that can cause unintended harms. The algorithmic fairness literature disambiguates between {\em representational} and {\em allocative} harms \cite{barocas-hardt-narayanan}.

{\em Representational harms} encompass the perpetuation of stereotypes or cultural norms that subject individuals to denigration. For example, airport body scanners routinely subject gender non-conforming passengers to public humiliation \cite{tsa}.

{\em Allocative harms} are when certain groups are denied access to a resource because of algorithmic bias. For example, a virtual try-on experience based on biased algorithms might exclude the precise people with non-normative bodies who are most in danger in traditional physical changing rooms \cite{changingroom}.

Finally, ignoring the existence of gender non-conforming individuals in our research (O3) creates an exclusionary environment for these members of our research community, contravening SIGGRAPH's goal to be \emph{a model of inclusion, equity, access and diversity}.

\section{Where do we go from here?}

Our analysis reveals that the common use of sex and gender in Computer Graphics can pepper our research with algorithmic bias. Our disambiguated study shows bias throughout the modeling process: algorithmic fairness cannot be an afterthought but must present at all stages of our research. We have focused on sex and gender, but hope our work broadens conversations about algorithmic fairness.%

Real-world constraints may make it unrealistic for specific research groups to mitigate some sources of bias, but potentially introduced biases should still be acknowledged. For example, none of the surveyed papers evaluated algorithmic fairness metrics (for a summary, see \cite{fairnesssurvey, fairnessmetrics}), nor discussed the potential harms of their treatment of sex and gender.

The issues raised by our survey often reveal \emph{scientific} limitations. If a method cannot model a class of humans, or a production system fails for a subsection of the population, these are fundamental \emph{technical} limitations, and should be discussed as such. Gender and sex can have a place in our research. It would be beneficial to report them among demographic statistics of datasets or user study participants (self-reported and non binary, in agreement with the scientific consensus) to
safeguard against the “male default” that plagues the sciences. In most cases where we observed sex or gender being used as features or targets, they should have been replaced by other, more accurate, variables. Finding these omitted variables and disaggregating the attributes that have been traditionally crammed into sex and gender constitute important open research problems.

Our proposed break with tradition requires effort, and difficult conversations. These are challenges worth facing if we want scientific advances to produce a fairer, more inclusive future.

\bibliographystyle{ACM-Reference-Format}
\bibliography{references.bib}


\begin{thebibliography}{22}


\ifx \showCODEN    \undefined \def \showCODEN     #1{\unskip}     \fi
\ifx \showDOI      \undefined \def \showDOI       #1{#1}\fi
\ifx \showISBNx    \undefined \def \showISBNx     #1{\unskip}     \fi
\ifx \showISBNxiii \undefined \def \showISBNxiii  #1{\unskip}     \fi
\ifx \showISSN     \undefined \def \showISSN      #1{\unskip}     \fi
\ifx \showLCCN     \undefined \def \showLCCN      #1{\unskip}     \fi
\ifx \shownote     \undefined \def \shownote      #1{#1}          \fi
\ifx \showarticletitle \undefined \def \showarticletitle #1{#1}   \fi
\ifx \showURL      \undefined \def \showURL       {\relax}        \fi
\providecommand\bibfield[2]{#2}
\providecommand\bibinfo[2]{#2}
\providecommand\natexlab[1]{#1}
\providecommand\showeprint[2][]{arXiv:#2}

\bibitem[\protect\citeauthoryear{Barocas, Hardt, and Narayanan}{Barocas
  et~al\mbox{.}}{2019}]%
        {barocas-hardt-narayanan}
\bibfield{author}{\bibinfo{person}{S. Barocas}, \bibinfo{person}{M. Hardt},
  {and} \bibinfo{person}{A. Narayanan}.} \bibinfo{year}{2019}\natexlab{}.
\newblock \bibinfo{booktitle}{\emph{Fairness and Machine Learning}}.
\newblock


\bibitem[\protect\citeauthoryear{Beauchamp}{Beauchamp}{2019}]%
        {tsa}
\bibfield{author}{\bibinfo{person}{T. Beauchamp}.}
  \bibinfo{year}{2019}\natexlab{}.
\newblock \bibinfo{booktitle}{\emph{Going Stealth: Transgender Politics and
  U.S. Surveillance Practices}}.
\newblock


\bibitem[\protect\citeauthoryear{Behzadi}{Behzadi}{2021}]%
        {cars}
\bibfield{author}{\bibinfo{person}{Y. Behzadi}.}
  \bibinfo{year}{2021}\natexlab{}.
\newblock \bibinfo{title}{Synthetic data to play a real role in enabling ADAS
  and autonomy}.
\newblock
\newblock


\bibitem[\protect\citeauthoryear{Blackless, Charuvastra, Derryck,
  Fausto-Sterling, Lauzanne, and Lee}{Blackless et~al\mbox{.}}{2000}]%
        {blackless2000sexually}
\bibfield{author}{\bibinfo{person}{M. Blackless}, \bibinfo{person}{A.
  Charuvastra}, \bibinfo{person}{A. Derryck}, \bibinfo{person}{A.
  Fausto-Sterling}, \bibinfo{person}{K. Lauzanne}, {and} \bibinfo{person}{E.
  Lee}.} \bibinfo{year}{2000}\natexlab{}.
\newblock \showarticletitle{How sexually dimorphic are we?}
\newblock \bibinfo{journal}{\emph{Am. J. Hum. Biol.}} \bibinfo{volume}{12},
  \bibinfo{number}{2} (\bibinfo{year}{2000}).
\newblock


\bibitem[\protect\citeauthoryear{Brewer}{Brewer}{2020}]%
        {dhs}
\bibfield{author}{\bibinfo{person}{T. Brewer}.}
  \bibinfo{year}{2020}\natexlab{}.
\newblock \showarticletitle{DHS Awards \$1 Million to Support Machine Learning
  Development for Airport Security}.
\newblock \bibinfo{journal}{\emph{Synthetik Applied Technologies Blog}}
  (\bibinfo{year}{2020}).
\newblock


\bibitem[\protect\citeauthoryear{Butler}{Butler}{2003}]%
        {butler2003gender}
\bibfield{author}{\bibinfo{person}{Judith Butler}.}
  \bibinfo{year}{2003}\natexlab{}.
\newblock \showarticletitle{Gender trouble}.
\newblock \bibinfo{journal}{\emph{Continental feminism reader}}
  (\bibinfo{year}{2003}), \bibinfo{pages}{29--56}.
\newblock


\bibitem[\protect\citeauthoryear{Chen, Lu, Chen, Williamson, and Mahmood}{Chen
  et~al\mbox{.}}{2021}]%
        {chen2021synthetic}
\bibfield{author}{\bibinfo{person}{R.~J. Chen}, \bibinfo{person}{M.~Y. Lu},
  \bibinfo{person}{T.~Y. Chen}, \bibinfo{person}{D.~FK Williamson}, {and}
  \bibinfo{person}{F. Mahmood}.} \bibinfo{year}{2021}\natexlab{}.
\newblock \showarticletitle{Synthetic data in machine learning for medicine and
  healthcare}.
\newblock \bibinfo{journal}{\emph{Nat. Biom.}} (\bibinfo{year}{2021}),
  \bibinfo{pages}{1--5}.
\newblock


\bibitem[\protect\citeauthoryear{Clarke}{Clarke}{2005}]%
        {clarke2005phantom}
\bibfield{author}{\bibinfo{person}{K.~A Clarke}.}
  \bibinfo{year}{2005}\natexlab{}.
\newblock \showarticletitle{The phantom menace: Omitted variable bias in
  econometric research}.
\newblock \bibinfo{journal}{\emph{Conflict management and peace science}}
  \bibinfo{volume}{22}, \bibinfo{number}{4} (\bibinfo{year}{2005}),
  \bibinfo{pages}{341--352}.
\newblock


\bibitem[\protect\citeauthoryear{Fausto-Sterling}{Fausto-Sterling}{2012}]%
        {fausto2012sex}
\bibfield{author}{\bibinfo{person}{A. Fausto-Sterling}.}
  \bibinfo{year}{2012}\natexlab{}.
\newblock \bibinfo{booktitle}{\emph{Sex/gender: Biology in a social world}}.
\newblock \bibinfo{publisher}{Routledge}.
\newblock


\bibitem[\protect\citeauthoryear{Friedman and Nissenbaum}{Friedman and
  Nissenbaum}{1996}]%
        {FriedmanAndNissenbaum}
\bibfield{author}{\bibinfo{person}{B. Friedman} {and} \bibinfo{person}{H.
  Nissenbaum}.} \bibinfo{year}{1996}\natexlab{}.
\newblock \showarticletitle{Bias in Computer Systems}.
\newblock \bibinfo{journal}{\emph{ACM Trans. Inf. Syst.}} \bibinfo{volume}{14},
  \bibinfo{number}{3} (\bibinfo{date}{jul} \bibinfo{year}{1996}),
  \bibinfo{pages}{330–347}.
\newblock
\showISSN{1046-8188}
\urldef\tempurl%
\url{https://doi.org/10.1145/230538.230561}
\showDOI{\tempurl}


\bibitem[\protect\citeauthoryear{{Google}}{{Google}}{2020}]%
        {googlegender}
\bibfield{author}{\bibinfo{person}{{Google}}.} \bibinfo{year}{2020}\natexlab{}.
\newblock \bibinfo{title}{{Ethics in Action: Removing Gender Labels from
  Cloud's Vision API}}.
\newblock
\newblock


\bibitem[\protect\citeauthoryear{Keyes}{Keyes}{2018}]%
        {keyes2018misgendering}
\bibfield{author}{\bibinfo{person}{O. Keyes}.} \bibinfo{year}{2018}\natexlab{}.
\newblock \showarticletitle{The misgendering machines: Trans/HCI implications
  of automatic gender recognition}.
\newblock \bibinfo{journal}{\emph{Proceedings of the ACM on human-computer
  interaction}} (\bibinfo{year}{2018}).
\newblock


\bibitem[\protect\citeauthoryear{Lorber}{Lorber}{1994}]%
        {lorber1994paradoxes}
\bibfield{author}{\bibinfo{person}{J. Lorber}.}
  \bibinfo{year}{1994}\natexlab{}.
\newblock \bibinfo{booktitle}{\emph{Paradoxes of gender}}.
\newblock \bibinfo{publisher}{Yale University Press}.
\newblock


\bibitem[\protect\citeauthoryear{Mehrabi, Morstatter, Saxena, Lerman, and
  Galstyan}{Mehrabi et~al\mbox{.}}{2021}]%
        {fairnesssurvey}
\bibfield{author}{\bibinfo{person}{N. Mehrabi}, \bibinfo{person}{F.
  Morstatter}, \bibinfo{person}{N. Saxena}, \bibinfo{person}{K. Lerman}, {and}
  \bibinfo{person}{A. Galstyan}.} \bibinfo{year}{2021}\natexlab{}.
\newblock \showarticletitle{A Survey on Bias and Fairness in Machine Learning}.
\newblock \bibinfo{journal}{\emph{ACM Comput. Surv.}} \bibinfo{volume}{54},
  \bibinfo{number}{6} (\bibinfo{year}{2021}).
\newblock


\bibitem[\protect\citeauthoryear{Money and Ehrhardt}{Money and
  Ehrhardt}{1972}]%
        {money1972man}
\bibfield{author}{\bibinfo{person}{J. Money} {and} \bibinfo{person}{A.
  Ehrhardt}.} \bibinfo{year}{1972}\natexlab{}.
\newblock \showarticletitle{Man and woman, boy and girl: Differentiation and
  dimorphism of gender identity from conception to maturity.}
\newblock  (\bibinfo{year}{1972}).
\newblock


\bibitem[\protect\citeauthoryear{Nature~Editors}{Nature~Editors}{2018}]%
        {pmid30377332}
\bibfield{author}{\bibinfo{person}{The Nature~Editors}.}
  \bibinfo{year}{2018}\natexlab{}.
\newblock \bibinfo{title}{{{U}{S} proposal for defining gender has no basis in
  science}}.
\newblock
\newblock


\bibitem[\protect\citeauthoryear{Olteanu, Castillo, Diaz, and
  K{\i}c{\i}man}{Olteanu et~al\mbox{.}}{2019}]%
        {olteanu2019social}
\bibfield{author}{\bibinfo{person}{A. Olteanu}, \bibinfo{person}{C. Castillo},
  \bibinfo{person}{F. Diaz}, {and} \bibinfo{person}{E. K{\i}c{\i}man}.}
  \bibinfo{year}{2019}\natexlab{}.
\newblock \showarticletitle{Social data: Biases, methodological pitfalls, and
  ethical boundaries}.
\newblock \bibinfo{journal}{\emph{Frontiers in Big Data}}  \bibinfo{volume}{2}
  (\bibinfo{year}{2019}), \bibinfo{pages}{13}.
\newblock


\bibitem[\protect\citeauthoryear{Pessach and Shmueli}{Pessach and
  Shmueli}{2020}]%
        {fairnessmetrics}
\bibfield{author}{\bibinfo{person}{D. Pessach} {and} \bibinfo{person}{E.
  Shmueli}.} \bibinfo{year}{2020}\natexlab{}.
\newblock \showarticletitle{Algorithmic Fairness}.
\newblock  (\bibinfo{year}{2020}).
\newblock
\showeprint[arXiv]{2001.09784}


\bibitem[\protect\citeauthoryear{Silver}{Silver}{2017}]%
        {changingroom}
\bibfield{author}{\bibinfo{person}{L. Silver}.}
  \bibinfo{year}{2017}\natexlab{}.
\newblock \showarticletitle{Topshop Refused To Let A Trans Person Into An
  All-Gender Changing Room}.
\newblock \bibinfo{journal}{\emph{BuzzFeed News}} (\bibinfo{year}{2017}).
\newblock


\bibitem[\protect\citeauthoryear{Suresh and Guttag}{Suresh and Guttag}{2021}]%
        {Suresh2021}
\bibfield{author}{\bibinfo{person}{H. Suresh} {and} \bibinfo{person}{J.
  Guttag}.} \bibinfo{year}{2021}\natexlab{}.
\newblock \showarticletitle{A Framework for Understanding Sources of Harm
  throughout the Machine Learning Life Cycle}.
\newblock \bibinfo{journal}{\emph{EAAMO}} (\bibinfo{date}{Oct}
  \bibinfo{year}{2021}).
\newblock


\bibitem[\protect\citeauthoryear{UNHCHR}{UNHCHR}{2015}]%
        {un2015report}
\bibfield{author}{\bibinfo{person}{UNHCHR}.} \bibinfo{year}{2015}\natexlab{}.
\newblock \showarticletitle{Discrimination and violence against individuals
  based on their sexual orientation and gender identity}.
\newblock  (\bibinfo{year}{2015}).
\newblock


\bibitem[\protect\citeauthoryear{{Unreal Engine}}{{Unreal Engine}}{2021}]%
        {metahuman}
\bibfield{author}{\bibinfo{person}{{Unreal Engine}}.}
  \bibinfo{year}{2021}\natexlab{}.
\newblock \bibinfo{title}{{Digital Humans | Metahuman Creator}}.
\newblock
\newblock


\end{thebibliography}

\end{document}